\newcommand{\bm}[1]{\mbox{\boldmath$#1$}}
\documentclass[twocolumn,showpacs,preprintnumbers,amsmath,amssymb]{revtex4}

\usepackage{graphicx}
\usepackage{dcolumn}
\usepackage{bm}

\begin{document}

\title{Phase diagram of a dilute ferromagnet model 
with antiferromagnetic next-nearest-neighbor interactions }

\author{ S. Niidera} 
\author{ S. Abiko}
 \altaffiliation[Present address: ]{FANUC LTD, Shibokusa 3580, Oshino-mura, 
          Yamanashi, Japan.} 
\author{ F. Matsubara} 
\email{fumi@camp.apph.tohoku.ac.jp}

\affiliation{Department of Applied Physics, Tohoku University, Sendai 980-8579,
Japan\\}

\date{ \today }

\begin{abstract}

We have studied the spin ordering of a dilute classical Heisenberg model 
with spin concentration $x$, and 
with ferromagnetic nearest-neighbor interaction $J_1$ and antiferromagnetic 
next-nearest-neighbor interaction $J_2$. 
Magnetic phases at absolute zero temperature $T = 0$ are determined 
examining the stiffness of the ground state, 
and those at finite temperatures $T \neq 0$ are determined 
calculating the Binder parameter $g_L$ and 
the spin correlation length $\xi_L$. 
Three ordered phases appear in the $x-T$ phase diagram: 
(i) the ferromagnetic (FM) phase; (ii) the spin glass (SG) phase; 
and (iii) the mixed (M) phase of the FM and the SG. 
Near below the ferromagnetic threshold $x_{\rm F}$, a reentrant 
SG transition occurs. 
That is, as the temperature is decreased from a high temperature, 
the FM phase, the M phase and the SG phase appear successively. 
The magnetization which grows in the FM phase disappears in the SG phase. 
The SG phase is suggested to be characterized by ferromagnetic clusters. 
We conclude, hence, that this model could reproduce experimental phase 
diagrams of dilute ferromagnets Fe$_x$Au$_{1-x}$ and Eu$_x$Sr$_{1-x}$S.

\end{abstract}


\maketitle


\section{Introduction}

Prototypes of spin glass (SG) are ferromagnetic dilute alloys such 
as Fe$_x$Au$_{1-x}$\cite{Coles}, Eu$_x$Sr$_{1-x}$S\cite{Maletta1,Maletta2} 
and Fe$_x$Al$_{1-x}$\cite{Shull,Motoya}. 
Those alloys have a common phase diagram as schematically shown in Fig. 1. 
It is shared with the ferromagnetic (FM) phase at higher spin 
concentrations and the SG phase at lower spin concentrations, 
together with the paramagnetic (PM) phase at high temperatures. 
A notable point is that {\it a reentrant spin glass (RSG) transition} 
occurs at the phase boundary between the FM phase and the SG phase. 
That is, as the temperature is decreased from a high temperature, 
the magnetization that grows in the FM phase vanishes at that phase 
boundary. The SG phase realized at lower temperatures is characterized 
by ferromagnetic clusters\cite{Coles,Maletta1,Maletta2,Motoya}. 
A similar phase diagram has also been reported for amorphous alloys 
$(T_{1-x}T'_x)_{75}$B$_6$Al$_3$ with $T$ = Fe or Co and $T'$ = Mn 
or Ni\cite{Yeshurun}. 
It is believed that the phase diagram of Fig. 1 arises from the competition 
between ferromagnetic and antiferromagnetic interactions. 
For example, in Fe$_x$Au$_{1-x}$, the spins are coupled via the long-range 
oscillatory Ruderman-Kittel-Kasuya-Yoshida (RKKY) interaction. 
Also, in Eu$_x$Sr$_{1-x}$S, the Heisenberg spins of $S = 7/2$ are coupled 
via short-range ferromagnetic nearest-neighbor exchange interaction 
and antiferromagnetic next-nearest-neighbor interaction\cite{EuSrS}. 
Nevertheless, the phase diagrams of the dilute alloys have not yet 
been understood theoretically. 
Several models have been proposed for explaining the RSG 
transition\cite{Saslow, Gingras1, Hertz}. 
However, no realistic model has been revealed that reproduces 
it\cite{Reger&Young,Gingras2,Morishita}.
Our primary question is, then, whether the experimental phase diagrams 
with the RSG transition are reproducible using a simple dilute model 
with competing ferromagnetic and antiferromagnetic interactions. 

\begin{figure}[bbb]
\vspace{-0.2cm}
\includegraphics[width=3.5cm,clip]{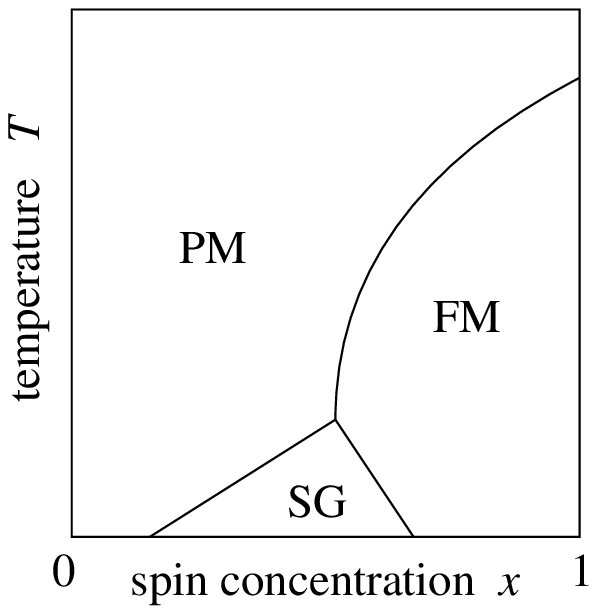}
\vspace{-0.2cm}
\caption{\label{fig:0}
A schematic phase diagram of a ferromagnetic dilute alloy. 
}
\end{figure}

This study elucidates a dilute Heisenberg model with competing 
short-range ferromagnetic nearest-neighbor exchange interaction $J_1$ and 
antiferromagnetic next-nearest-neighbor interaction $J_2$. 
This model was examined nearly 30 years ago using a computer simulation 
technique\cite{Binder} at rather high spin concentrations 
and the phase boundary between the PM phase and the FM phase 
was obtained. 
However, the SG transition and the RSG transition have not yet been examined. 
Recent explosive advances in computer power have enabled us to perform larger 
scale computer simulations. 
Using them, we reexamine the spin ordering of the model for both $T = 0$ and 
$T \neq 0$ in a wide-spin concentration range. 
Results indicate that the model reproduces qualitatively the experimental 
phase diagrams. 
In particular, we show that the model reproduces the RSG transition. 
A brief report of this result was given in Ref. 15. 


The paper is organized as follows. In Sec. II, we present the model. 
In Sec. III, the ground state properties are discussed. We will determine 
threshold $x_{\rm F}$, above which the ground state magnetization remains 
finite. Then we examine the stabilities of the FM phase and the SG phase 
calculating excess energies that are obtained by twisting the ground state 
spin structure. 
Section IV presents Monte Carlo simulation results. 
We will give both the phase boundaries between the PM phase and 
the FM phase and between the PM phase and the SG phase. 
Immediately below $x = x_{\rm F}$, we find the RSG transition. 
Section V is devoted to our presentation of important conclusions.

\section{Model}

We start with a dilute Heisenberg model with competing nearest-neighbor and 
next-nearest-neighbor exchange interactions described by the Hamiltonian: 
\begin{eqnarray} 
 H = &-& \sum_{\langle ij \rangle}^{nn}J_1x_ix_j\bm{S}_{i}\cdot\bm{S}_{j} 
   + \sum_{\langle kl \rangle}^{nnn}J_2x_kx_l\bm{S}_{k}\cdot\bm{S}_{l}, 
\end{eqnarray} 
where $\bm{S}_{i}$ is the classical Heisenberg spin of $|\bm{S}_{i}| = 1$; 
$J_1 (> 0)$ and $J_2 (> 0)$ respectively represent the nearest-neighbor 
and the next-nearest-neighbor exchange interactions; and $x_i = 1$ and 0 when the 
lattice site $i$ is occupied respectively by a magnetic and non-magnetic atom. 
The average number of $x (\equiv \langle x_i \rangle)$ is the concentration 
of a magnetic atom. 
Note that an experimental realization of this model is 
Eu$_x$Sr$_{1-x}$S\cite{EuSrS}, in which magnetic atoms (Eu) are located on 
the fcc lattice sites. 
Here, for simplicity, we consider the model on a simple cubic lattice with 
$J_2 = 0.2J_1$\cite{Model}. 


\section{Magnetic Phase at $T = 0$}

We consider the magnetic phase at $T = 0$. Our strategy is as follows. 
First we consider the ground state of the model on finite lattices 
for various spin concentrations $x$. 
Examining the size dependence of magnetization $M$, 
we determine the spin concentration $x_{\rm F}$ above which the magnetization 
will take a finite, non-vanishing value for $L \rightarrow \infty$. 
Then we examine the stability of the ground state by calculating 
twisting energies. 
We apply a hybrid genetic algorithm (HGA)\cite{GA} for searching for 
the ground state.


\subsection{Magnetization $M$ at $T = 0$}

We treat lattices of $L \times L \times L$ with 
periodic boundary conditions. 
The ground state magnetizations $\bm{M}_L^{\rm G} ( \equiv 
\sum_{i}x_i\bm{S}_i)$ are calculated for individual samples 
and averaged over the samples. That is, $M = [|\bm{M}_L^{\rm G}|]$, 
where $[ \cdots ]$ represents a sample average.
Numbers $N_s$ of samples with different spin distributions are 
$N_s = 1000$ for $L \leq 8$, $N_s = 500$ for $10 \leq L \leq 14$, and 
$N_s = 64$ for $L \geq 16$. 
We apply the HGA with the number $N_p$ of parents of $N_p = 16$ 
for $L \leq 8$, $N_p = 64$ for $L = 10$, $N_p = 128$ for $L = 12$, 
$\dots$, and $N_p = 512$ for $L \geq 16$. 


\begin{figure}[tb]
\includegraphics[width=6.5cm,clip]{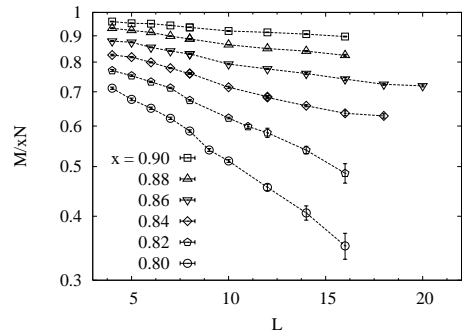}
\vspace{-0.4cm}
\caption{\label{fig:1}
Ground state magnetizations $M$ in $L \times L \times L$ lattices 
for various spin concentrations $x$. 
}
\end{figure}

\begin{figure}[tb]
\includegraphics[width=6.5cm,clip]{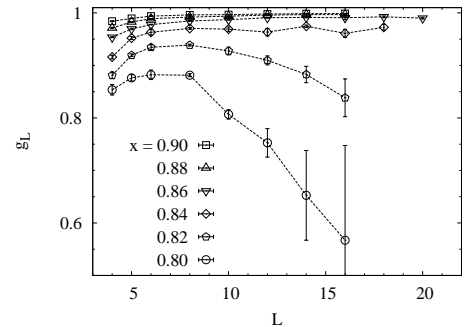}
\vspace{-0.4cm}
\caption{\label{fig:2}
Binder parameter $g_L$ at $T = 0$ in $L \times L \times L$ lattices 
for various spin concentrations $x$. 
}
\end{figure}

Figure 2 portrays plots of magnetization $M$ as a function of $L$ for 
various spin concentrations $x$. 
A considerable difference is apparent in the $L$-dependence of $M$ 
between $x \leq 0.82$ and $x \geq 0.84$. 
For $x \leq 0.82$, 
as $L$ increases, $M$ decreases exponentially revealing that 
$M \rightarrow 0$ for $L \rightarrow \infty$. On the other hand, for 
$x \geq 0.84$, $M$ decreases rather slowly, suggesting that $M$ remains finite 
for $L \rightarrow \infty$.

To examine the above suggestion, we calculate the Binder parameter 
$g_L$\cite{BinderP} defined as
\begin{eqnarray} 
g_L = (5 - 3\frac{[|\bm{M}_L^{\rm G}|^4]}{[|\bm{M}_L^{\rm G}|^2]^2})/2. 
\end{eqnarray} 
When the sample dependence of $\bm{M}_L^{\rm G}$ vanishes for 
$L \rightarrow \infty$, $g_L$ increases with $L$ and becomes unity. 
That is, if the system has its magnetization inherent in the system, 
$g_L$ increases with $L$. 
On the other hand, $g_L \rightarrow 0$ for $L \rightarrow \infty$ 
when $\bm{M}_L^{\rm G}$ tends to scatter according to a Gaussian 
distribution. 
Figure 3 represents the $L$-dependence of $g_L$ for various $x$. 
For $x \leq 0.82$, as $L$ increases, $g_L$ increases and subsequently 
becomes maximum at $L \sim 8$, decreasing thereafter. 
This fact reveals that the FM phase is absent 
for $x \leq 0.82$. 
For $x \geq 0.84$, a decrease is not apparent. 
In particular, $g_L$ for $x \geq 0.86$ increases gradually toward 1, 
indicating that the FM phase occurs for $L \rightarrow \infty$. 
We suggest, hence, the threshold of the FM phase of 
$x_{\rm F} = 0.84 \pm 0.02$ at $T = 0$.


\subsection{Stiffness of the ground state}

The next question is, for $x > x_{\rm F}$, whether or not the FM 
phase is stable against a weak perturbation. 
Also, for $x < x_{\rm F}$, whether or not some frozen spin structure occurs. 
To consider these problems, we examine the stiffness of 
the ground state\cite{Endoh1,Endoh2}. 


We briefly present the method\cite{Endoh1}. 
We consider the system on a cubic lattice with $L \times L \times (L+1)$ 
lattice sites in which the $z$-direction is chosen as one for $(L+1)$ 
lattice sites. 
That is, the lattice is composed of $(L+1)$ layers with 
$L \times L$ lattice sites. 
Periodic boundary conditions are applied for every layer and 
an open boundary condition to the $z$-direction. 
Therefore, the lattice has two opposite surfaces: 
$\Omega_1$ and $\Omega_{L+1}$. 
We call this system as the reference system. First, we determine 
the ground state of the reference system. 
We denote the ground state spin configuration on the $l$th layer 
as $\{ \bm{S}_{l,i}\} \ (l = 1$ -- $(L+1) )$ and the ground state 
energy as $E_L^{\rm G}$. 
Then we add a distortion inside the system in such a manner that, under 
a condition that $\{ \bm{S}_{1,i}\}$ are fixed, $\{ \bm{S}_{L+1,i}\}$ 
are rotated by the same angle $\phi$ around some common axis. 
We also call this system a twisted system. 
The minimum energy $E_L(\phi)$ of the twisted system is always higher 
than $E_L^{\rm G}$. 
The excess energy $\Delta E_L(\phi) (\equiv E_L(\phi) - E_L^{\rm G})$ is 
the net energy that is added inside the lattice by this twist, because 
the surface energies of $\Omega_{1}$ and $\Omega_{L+1}$ are conserved. 
The stiffness exponent $\theta$ may be defined by the relation 
$\Delta E_L(\phi) \propto L^{\theta}$\cite{Comm_Endoh}.
If $\theta > 0$, the ground state spin configuration is stable 
against a small perturbation. That is, the ground state phase will occur 
at least at very low temperatures. 
On the other hand, if $\theta < 0$, the ground state phase is absent 
at any non-zero temperature.


To apply the above idea to our model, we must give special attention 
to the rotational axis for $\{ \bm{S}_{L+1,i}\}$ because 
the reference system has a non-vanishing magnetization ${\bm M}_L^G$. 
For the following arguments, we separate each spin ${\bm S}_{l,i}$ 
into parallel and perpendicular components:
\[ \left\{ 
\begin{array}{l}
{\bm S}_{l,i}^{\parallel} = ({\bm S}_{l,i}\cdot{\bm m}){\bm m} \\
{\bm S}_{l,i}^{\perp}     = ({\bm S}_{l,i} \times {\bm m}) \times {\bm m}, 
\end{array} \right.
\]
where ${\bm m} = {\bm M}_L^{\rm G}/|{\bm M}_L^{\rm G}|$. 
We consider two twisted systems. 
One is a system in which $\{ \bm{S}_{L+1,i}^{\perp}\}$ are rotated around 
the axis that is parallel to the magnetization ${\bm M}_L^{\rm G}$. 
We denote the minimum energy of this twisted system as $E_L^{\perp}(\phi)$. 
The other is a system in which $\{ \bm{S}_{L+1,i}\}$ are rotated around 
an axis that is perpendicular to ${\bm M}_L^{\rm G}$. 
We also denote the minimum energy of this twisted system as 
$E_L^{\parallel}(\phi)$. 
Note that, in this twisted system, $\{{\bm S}_{l,i}^{\parallel}\}$ mainly 
change, but $\{{\bm S}_{l,i}^{\perp}\}$ also change. 
Choices in the rotation axis are always possible in finite systems, 
even when $x < x_{\rm F}$ because a non-vanishing magnetization 
(${\bm M}_L^{\rm G} \neq 0$) exists in the Heisenberg model on a finite 
lattice. 
Of course the difference between $E_L^{\perp}(\phi)$ and 
$E_L^{\parallel}(\phi)$ will diminish for $L \rightarrow \infty$ 
in the range $x < x_{\rm F}$.
The excess energies $\Delta E_L^{\perp}(\phi)$ and 
$\Delta E_L^{\parallel}(\phi)$ in our model are given as 
\begin{eqnarray}
  \Delta E_L^{\perp}(\phi) &=& [E_L^{\perp}(\phi) - E_L^{\rm G}],  \\
  \Delta E_L^{\parallel}(\phi) &=& [E_L^{\parallel}(\phi) -E_L^{\rm G}],  
\end{eqnarray}
with $[\cdots]$ being the sample average.


We calculated $\Delta E_L^{\perp}(\phi)$ and $\Delta E_L^{\parallel}(\phi)$ 
for a common rotation angle of $\phi = \pi/2$ in lattices of $L \leq 14$. 
Numbers of the samples are $N_s \sim 1000$ for $L \leq 10$ and 
$N_s \sim 250$ for $L = 12$ and 14. 
Hereafter we simply describe $\Delta E_L^{\perp}(\pi/2)$ and 
$\Delta E_L^{\parallel}(\pi/2)$ respectively as $\Delta E_L^{\perp}$ 
and $\Delta E_L^{\parallel}$. 
Figures 4(a) and 4(b) respectively show lattice size dependences of 
$\Delta E_L^{\perp}$ and $\Delta E_L^{\parallel}$ for $x < x_{\rm F}$ 
and $x > x_{\rm F}$. 
We see that, for all $x$, $\Delta E_L^{\parallel} > \Delta E_L^{\perp}$ and 
both increase with $L$. 
When $x < x_{\rm F}$, as expected, the difference between $\Delta E_L^{\perp}$ 
and $\Delta E_L^{\parallel}$ diminishes as $L$ increases. 

\begin{figure}[tb]
\includegraphics[width=6.5cm,clip]{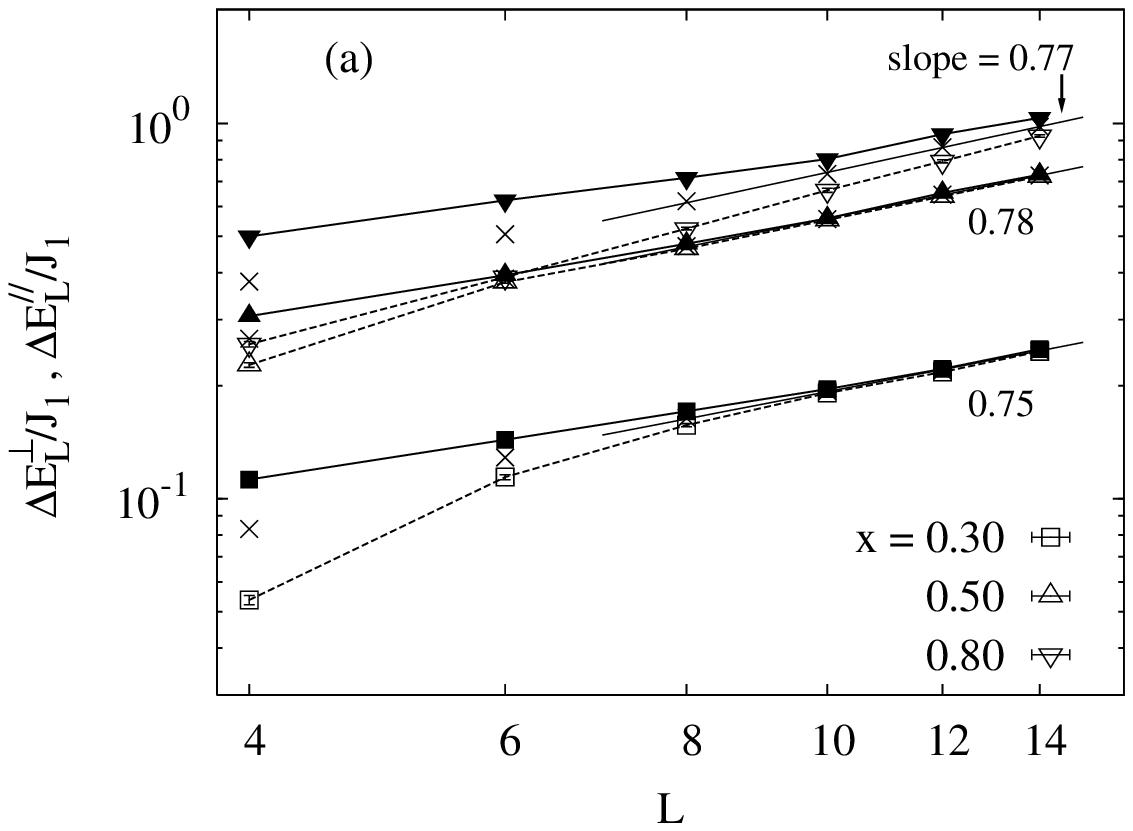}
\vspace{-0.2cm}
\includegraphics[width=6.5cm,clip]{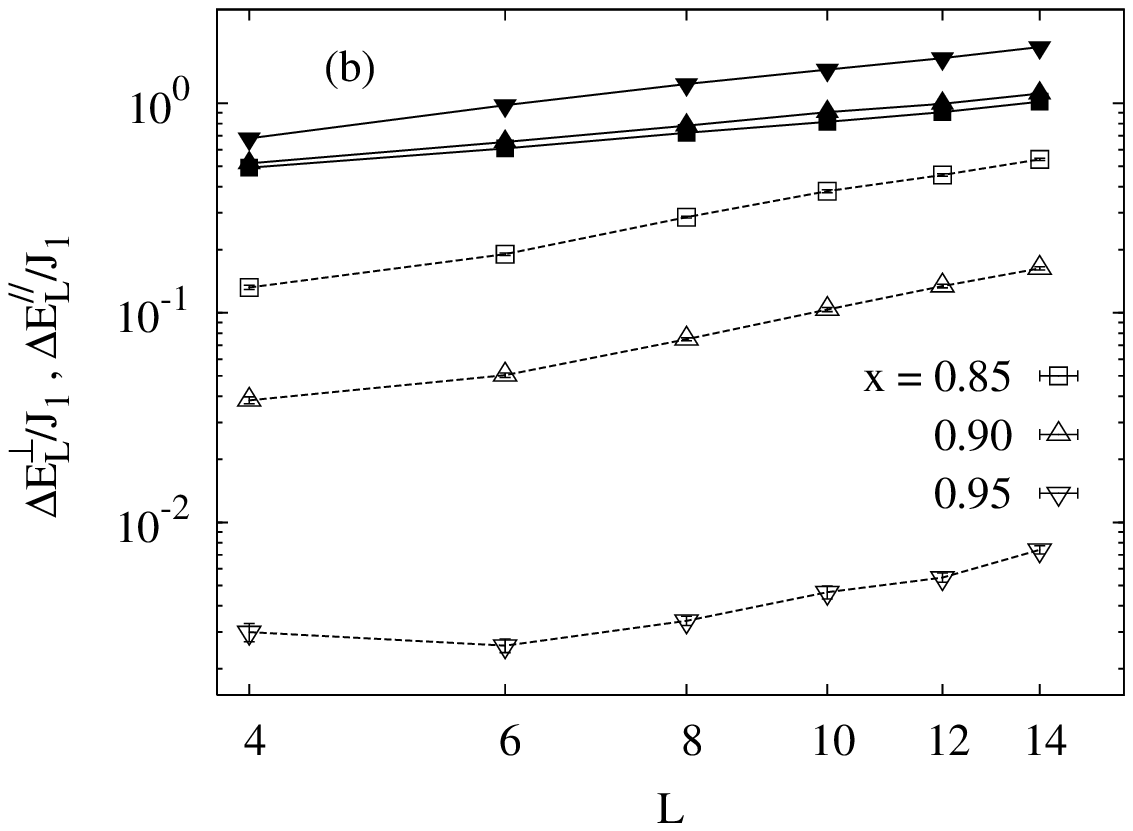}
\vspace{-0.2cm}
\caption{\label{fig:3}
Excess energies $\Delta E_L^{\perp}$ and $\Delta E_L^{\parallel}$ for 
$L \times L \times (L+1)$ lattices for various spin concentrations: 
(a) $x < x_{\rm F}$ and (b) $x > x_{\rm F}$. 
Open symbols represent $\Delta E_L^{\perp}$ and filled symbols 
$\Delta E_L^{\parallel}$. 
Symbols $\times$ in (a) represent the averages of those values. 
}
\end{figure}

Now we discuss the stability of the spin configuration. 
First we consider the stability of $\{{\bm S}_{l,i}^{\parallel}\}$, i.e., 
the stability of the FM phase. 
In the pure FM case ($x = 1$), ${\bm S}_{l,i}^{\perp} = 0$ and 
$\Delta E_L^{\parallel}$ gives the net excess energy for the twist of 
the magnetization. 
This is not the same in the case of ${\bm S}_{l,i}^{\perp} \neq 0$. 
Because the twist in $\{{\bm S}_{l,i}^{\parallel}\}$ accompanies the change 
in $\{{\bm S}_{l,i}^{\perp}\}$, $\Delta E_L^{\parallel}$ does not give 
the net excess energy for the twist of $\{ {\bm S}_{l,i}^{\parallel}\}$. 
For that reason, we consider the difference $\Delta E_L^{\rm F}$ between 
the two excess energies: 
\begin{eqnarray}
  \Delta E_L^{\rm F} = \Delta E_L^{\parallel}-\Delta E_L^{\perp}. 
\end{eqnarray}
If $\Delta E_L^{\rm F} \rightarrow \infty$ for $L \rightarrow \infty$, 
the FM phase will be stable against a small perturbation. 
We define the stiffness exponent $\theta^{\rm F}$ of the FM 
phase as 
\begin{eqnarray}
  \Delta E_L^{\rm F} \propto L^{\theta^{\rm F}}. 
\end{eqnarray}
Figure 5 shows $\Delta E_L^{\rm F}$ for $x \geq 0.80$. 
We have $\theta^{\rm F} > 0$ for $x \geq 0.85$ and $\theta^{\rm F} < 0$ 
for $x = 0.80$. 
These facts show that, in fact, the FM phase is stable for 
$x > x_{\rm F} \sim 0.84$ at $T \sim 0$. 

\begin{figure}[tb]
\includegraphics[width=6.5cm,clip]{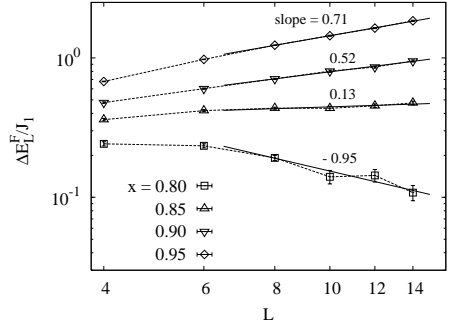}
\vspace{-0.2cm}
\caption{\label{fig:4}
Difference in the excess energy $\Delta E_L^F = \Delta E_L^{\parallel} - 
\Delta E_L^{\perp}$ for $L \times L \times (L+1)$ lattice for 
various spin concentrations $x$. 
}
\end{figure}

Next, we consider the stability of the transverse components 
$\{{\bm S}_{l,i}^{\perp}\}$. Hereafter we call the phase with 
$\{{\bm S}_{l,i}^{\perp} \neq 0\}$ a SG phase. For $x < x_{\rm F}$, 
we may examine the stiffness exponent $\theta^{\rm SG}$ using 
either $\Delta E_L^{\perp}$ or $\Delta E_L^{\parallel}$. 
Here we estimate its value using an average value of them. 
For $x > x_{\rm F}$, we examine it using $\Delta E_L^{\perp}$. 
In this range of $x$, meticulous care should be given to a strong finite 
size effect\cite{Comm_finite}.
We infer that this finite size effect for $x > x_{\rm F}$ is attributable to 
a gradual decrease in the magnetization ${\bm M}$ for finite $L$ (see Fig. 2). 
That is, the magnitude of the transverse component $|{\bm S}_{l,i}^{\perp}|$ 
will gradually increase with $L$, which will engender an additional 
increase of $\Delta E_L^{\perp}$ as $L$ increases. 
This increase of $|{\bm S}_{l,i}^{\perp}|$ will cease for 
$L \rightarrow \infty$. 
Consequently, we estimate the value of $\theta^{\rm SG}$ from the relations: 
\begin{eqnarray}
 (\Delta E_L^{\parallel}+\Delta E_L^{\perp})/2 &\propto& L^{\theta^{\rm SG}} 
 \ \ \ {\rm for} \ \ \ \ x < x_{\rm F}, \\
  \Delta E_L^{\perp}/|{\bm S}^{\perp}|^2 &\propto& L^{\theta^{\rm SG}}
 \ \ \ {\rm for} \ \ \ \ x > x_{\rm F},  
\end{eqnarray}
where $|{\bm S}^{\perp}|^2 = 1 - |{\bm M}/xN|^2$. 
Log-log plots of those quantities versus $L$ are presented in Fig. 4(a) 
for $x < x_{\rm F}$ and in Fig. 6 for $x > x_{\rm F}$. 
We estimate $\theta^{\rm SG}$ using data for $L \geq 8$ and present 
the results in the figures. 
Note that for $x > 0.90$, studies of bigger lattices will be necessary 
to obtain a reliable value of $\theta^{\rm SG}$ because $\Delta E_L^{\perp}$ 
for $L \lesssim 14$ is too small to examine the stiffness of 
$\{{\bm S}_{l,i}^{\perp}\}$.


Figure 7 shows stiffness exponents $\theta^{\rm F}$ and $\theta^{\rm SG}$ 
as functions of $x$. 
As $x$ increases, $\theta^{\rm SG}$ changes its sign from negative 
to positive at $x_{SG} = 0.175 \pm 0.025$. This value of $x_{\rm SG}$ is 
close to the percolation threshold of $x_{\rm p} \sim 0.137$\cite{Essam}. 
Above $x_{\rm SG}$, $\theta^{\rm SG}$ takes almost the same value of 
$\theta^{\rm SG} \sim 0.75$ up to $x \sim 0.9$. 
On the other hand, $\theta^{\rm F}$ changes its sign at $x_{\rm F} \sim 0.84$ 
and increases toward $\theta^{\rm F} = 1$ at $x = 1$. 
A notable point is that $\theta^{\rm SG} > 0$ for $x > x_{\rm F}$. 
That is, a mixed (M) phase of the ferromagnetism and the SG phase will occur 
for $x > x_{\rm F}$ at $T = 0$. 
We could not estimate another threshold of $x$ above which the purely 
FM phase is realized.

\begin{figure}[tb]
\includegraphics[width=6.5cm,clip]{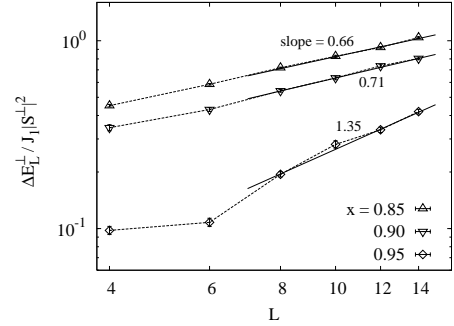}
\vspace{-0.2cm}
\caption{\label{fig:5}
The normalized excess energy $\Delta E_L^{\perp}/|{\bm S}^{\perp}|^2$ for 
$L \times L \times (L+1)$ lattices for various spin concentrations $x > x_{\rm F}$. }
\end{figure}

\begin{figure}[tb]
\includegraphics[width=6.5cm,clip]{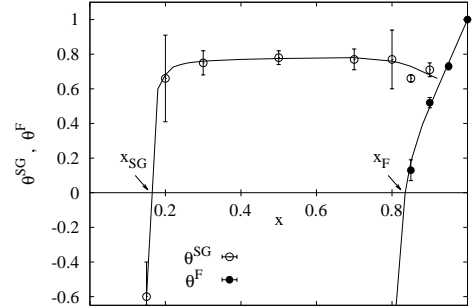}
\vspace{-0.2cm}
\caption{\label{fig:6}
Stiffness exponents $\theta^{\rm SG}$ and $\theta^{\rm F}$ for various spin 
concentrations $x$. Here, we remove $\theta^{\rm SG}$ at $x = 0.95$. 
}
\end{figure}


\section{Monte Carlo Simulation}

We next consider the magnetic phase at finite temperatures using 
the MC simulation technique. 
We make a MC simulation for $x \geq 0.20$. 
We treat lattices of $L \times L \times L \ (L= 8-48)$ with 
periodic boundary conditions. 
Simulation is performed using a conventional heat-bath MC method. 
The system is cooled gradually from a high temperature (cooling simulation). 
For larger lattices, $200 000$ MC steps (MCS) are allowed for 
relaxation; data of successive $200 000$ MCS are used to calculate 
average values. 
We will show later that these MCS are sufficient for studying 
equilibrium properties of the model at a temperature range 
within which the RSG behavior is found. 
Numbers $N_s$ of samples with different spin distributions are 
$N_s = 1000$ for $L \leq 16$, $N_s = 500$ for $L = 24$, 
$N_s = 200$ for $L = 32$, and $N_s = 80$ for $L = 48$. 
We measure the temperature in units of $J_1$ ($k_{\rm B} = 1$). 


\subsection{Thermal and magnetic properties}

We calculate the specific heat $C$ and magnetization $M$ given by 
\begin{eqnarray} 
 C &=& \frac{1}{T^2}([\langle E(s)^2 \rangle] - [\langle E(s) \rangle^2]),\\
 M &=& [\langle M(s) \rangle]. 
\end{eqnarray} 
Therein, $E(s)$ and $M(s) (\equiv |\sum_ix_i\bm{S}_i|)$ represent 
the energy and magnetization at the $s$th MC step, and $N$ is the 
number of the lattice sites. 
Here $\langle \cdots \rangle$ represents an MC average.

\begin{figure}[tb]
\includegraphics[width=7cm,clip]{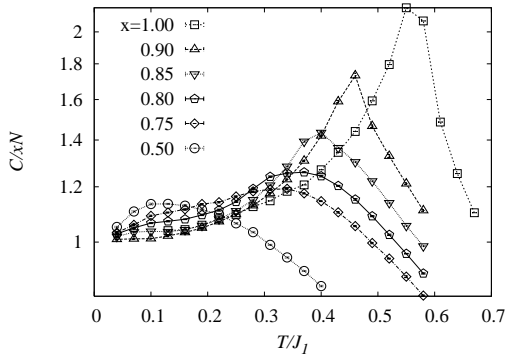}
\vspace{-0.2cm}
\caption{\label{fig:7}
Specific heats $C$ in the $32 \times 32 \times 32$ lattice 
for various spin concentrations $x$. }
\end{figure}

\begin{figure}[tb]
\includegraphics[width=7cm,clip]{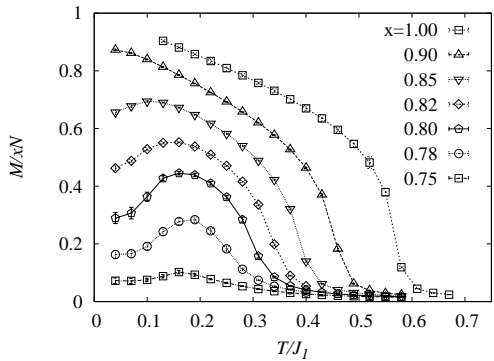}
\vspace{-0.2cm}
\caption{\label{fig:8}
Magnetizations $M$ in the $32 \times 32 \times 32$ lattice for 
various spin concentrations $x$. 
}
\end{figure}

Figure 8 shows the specific heat $C$ for various concentrations $x$. 
For $x \geq 0.90$, $C$ exhibits a sharp peak at a high temperature, 
revealing that a FM phase transition occurs at that temperature. 
As $x$ decreases, the peak broadens. 
On the other hand, at $x \sim 0.85$ a hump is apparent at a lower temperature; 
it grows with decreasing $x$. 
This fact implies that, for $x \lesssim 0.85$, another change in the spin 
structure occurs at a lower temperature. As $x$ decreases further, the broad 
peak at a higher temperature disappears and only a single broad peak 
is visible at a lower temperature. 


Figure 9 shows temperature dependencies of magnetization $M$ for various $x$. 
For $x = 1$, as the temperature decreases, $M$ increases rapidly 
below the temperature, revealing the occurrence of a FM phase. 
As $x$ decreases, $M$ exhibits an interesting phenomenon: 
in the range of $0.78 \lesssim x \lesssim 0.85$, $M$ once increases, 
reaches a maximum value, then decreases. 
We also perform a complementary simulation to examine this behavior of $M$. 
That is, starting with a random spin configuration at a low temperature, 
the system is heated gradually (heating simulation). 
Figure 10 shows temperature dependencies of $M$ for $x = 0.80$ 
in both cooling and heating simulations for various $L$. 
For $T \gtrsim 0.1J_1$, data of the two simulations almost coincide 
mutually, even for large $L$. 
We thereby infer that $M$ for $T \gtrsim 0.1J_1$ are of thermal 
equilibrium and the characteristic behavior of $M$ found here is an inherent 
property of the model. 
For $T < 0.1J_1$, a great difference in $M$ is apparent 
between the two simulations; estimation of the equilibrium value is difficult. 
We speculate, however, that the heating simulation gives a value of $M$ 
that is similar to that in the equilibrium state because the data in the 
heating simulation seem to concur with those obtained in the ground state. 


Figure 10 shows the remarkable lattice size dependence of $M$. 
For smaller $L$, as the temperature decreases, $M$ decreases slightly 
at very low temperatures. The decrease is enhanced as $L$ increases. 
Consequently, a strong size-dependence of $M$ is indicated for 
$T \lesssim 0.1J_1$. 
These facts suggest that $M$ for $L \rightarrow \infty$ disappears 
at low temperatures as well as at high temperatures. 
The next section presents an examination of this issue, calculating the 
Binder parameter.

\begin{figure}[tb]
\includegraphics[width=7.0cm,clip]{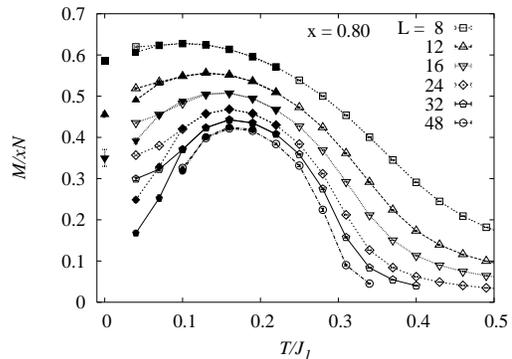}
\vspace{-0.4cm}
\caption{\label{fig:9}
Magnetizations $M$ for $x = 0.80$ in the $L\times L\times L$ lattice. 
Open symbols indicate $M$ in the cooling simulation and filled symbols 
indicate that in the heating simulation. Data at $T = 0$ indicate those 
in the ground state given in Fig. 2. 
}
\end{figure}


\subsection{Ferromagnetic phase transition}

The Binder parameter $g_L$ at finite temperatures is defined as
\begin{eqnarray} 
 g_L = (5 - 3\frac{[\langle M(s)^4\rangle]}{[\langle M(s)^2\rangle]^2})/2. 
\end{eqnarray} 
We calculate $g_L$ for various $x$. 
Figures 11(a)--11(d) show $g_L$'s for $x \sim x_{\rm F}$\cite{Comm_gL}. 
In fact, $g_L$ for $x < x_{\rm F}$ exhibits a novel temperature dependence. 
As the temperature is decreased from a high temperature, $g_L$ increases 
rapidly, becomes maximum, then decreases. 
In particular, we see in Fig. 11(b) for $x = 0.80$ $g_L$'s for different $L$ 
cross at two temperatures $T_{\rm C}$ and $T_{\rm R}$ ($< T_{\rm C}$). 
The cross at $T_{\rm C}/J_1 \sim 0.26$ is a usual one that is found 
in the FM phase transition. 
That is, for $T > T_{\rm C}$, $g_L$ for a larger size 
is smaller than that for a smaller size; for $T < T_{\rm C}$, 
this size dependence in $g_L$ is reversed. 
On the other hand, the cross at $T_{\rm R}$ is strange: 
for $T < T_{\rm R}$, $g_L$ for a larger size again becomes smaller than 
that for a smaller size. 
Interestingly, the cross for different $g_L$ occur at almost 
the same temperature of $T_{\rm R}/J_1 \sim 0.13$. 
These facts reveal that, as the temperature is decreased to 
below $T_{\rm R}$, the FM phase, which occurs below $T_{\rm C}$, 
disappears. Similar properties are apparent for $x =$ 0.79--0.82.

\begin{figure}[tb]
\begin{center}
\includegraphics[width=6.0cm,clip]{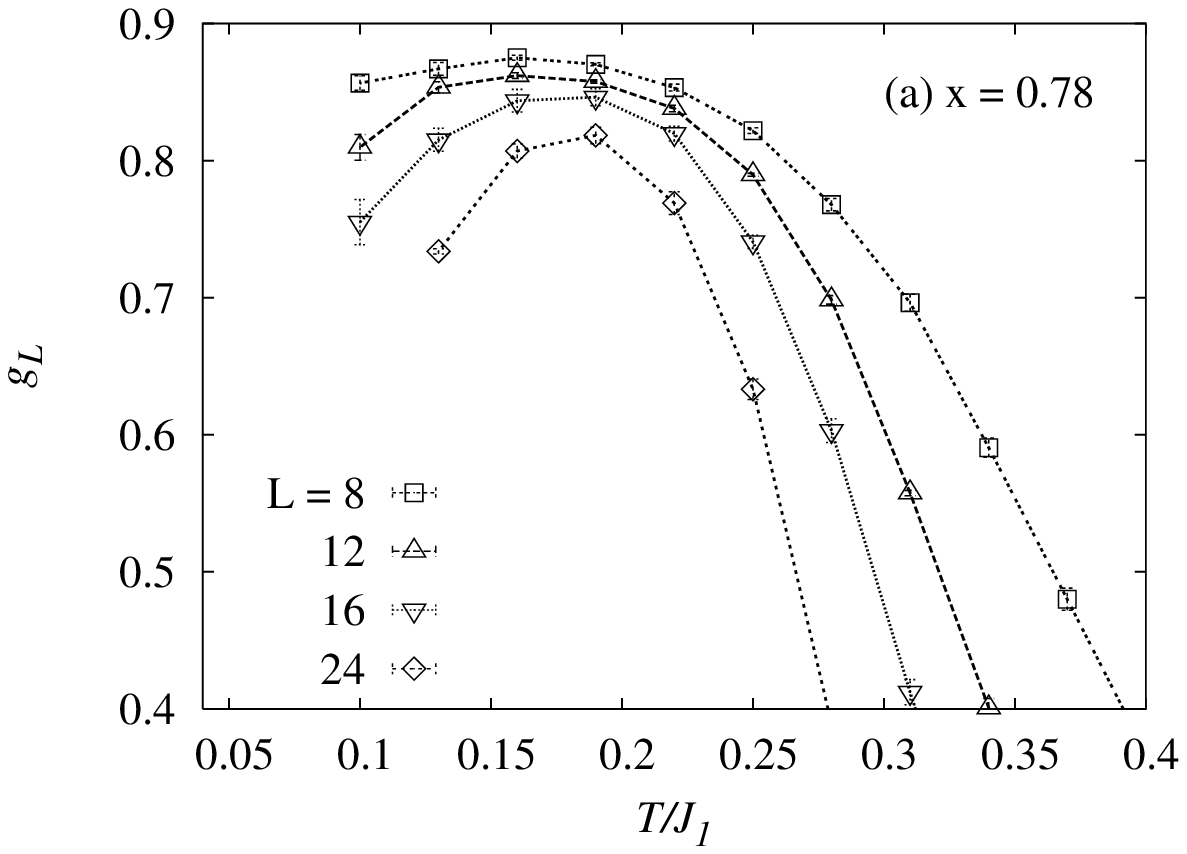}\\
\vspace{-0.2cm}
\includegraphics[width=6.0cm,clip]{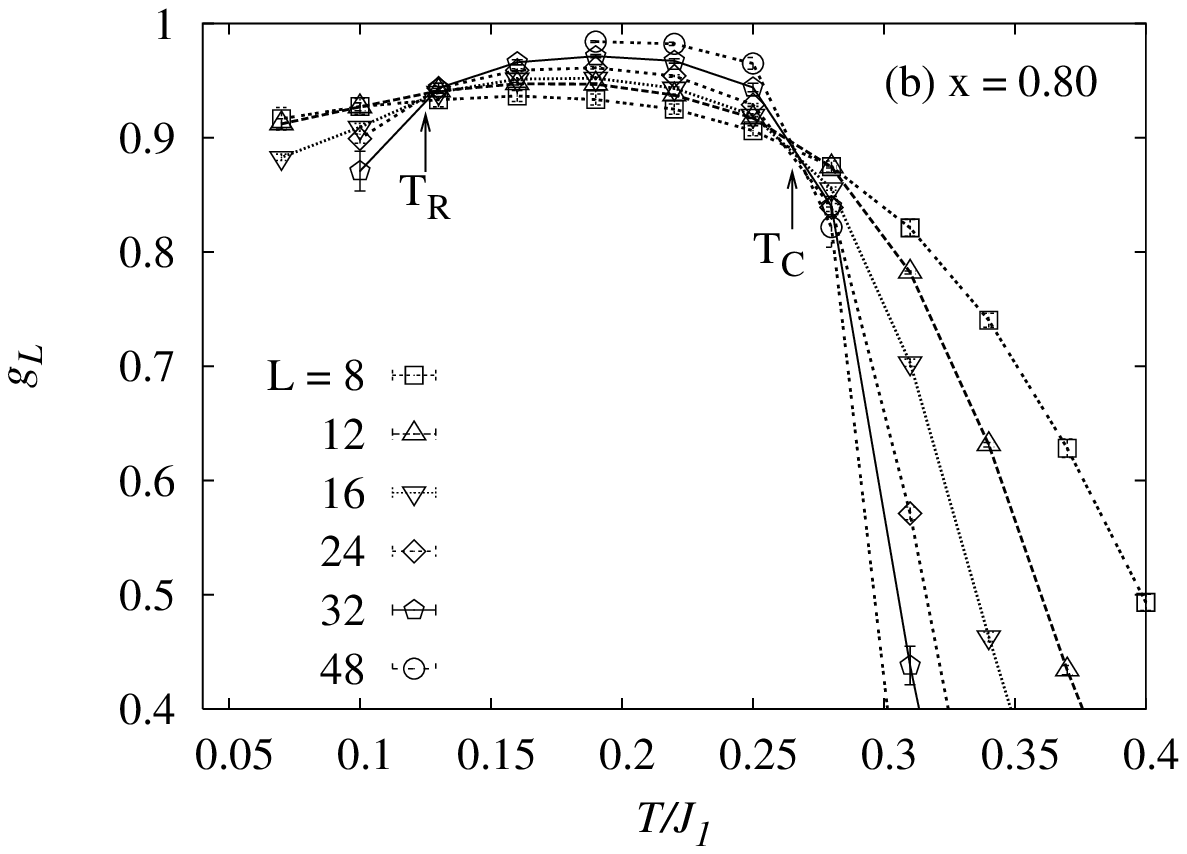}\\
\vspace{-0.2cm}
\includegraphics[width=6.0cm,clip]{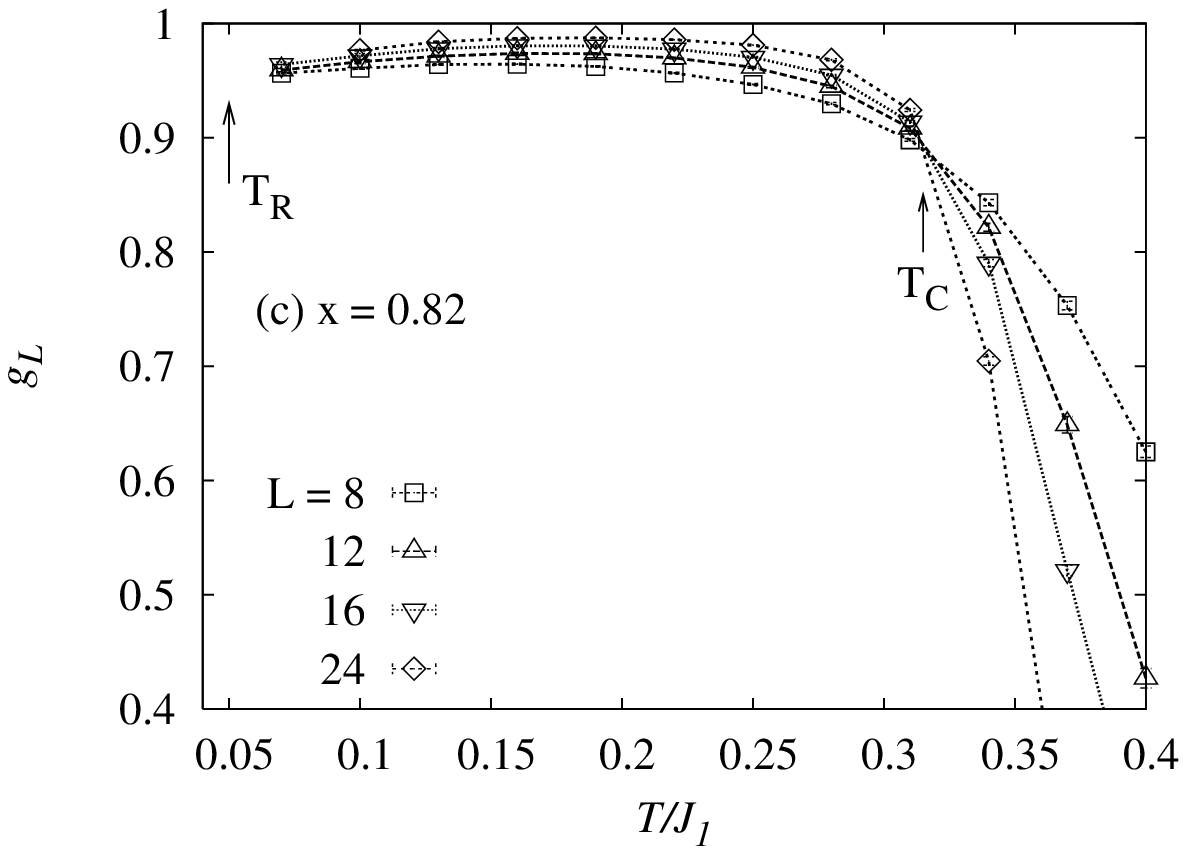}\\
\vspace{-0.2cm}
\includegraphics[width=6.0cm,clip]{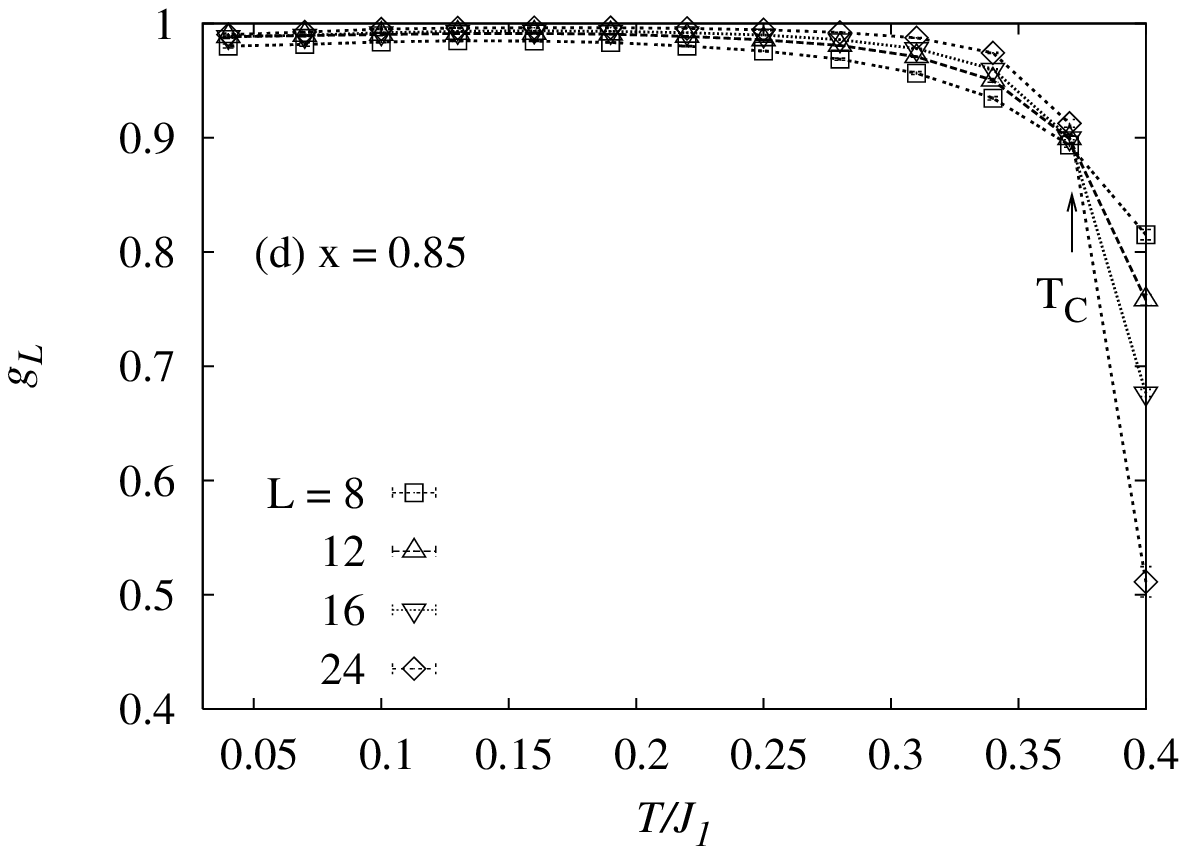}\\
\end{center}
\vspace{-0.4cm}
\caption{\label{fig:10}
Binder parameters $g_L$ for various $x$. The $T_{\rm R}$ for $x = 0.82$ 
was estimated by extrapolations of data obtained at higher temperatures. 
}
\end{figure}

\subsection{Spin glass phase transition}

\begin{figure}[tb]
\begin{center}
\includegraphics[width=7.0cm,clip]{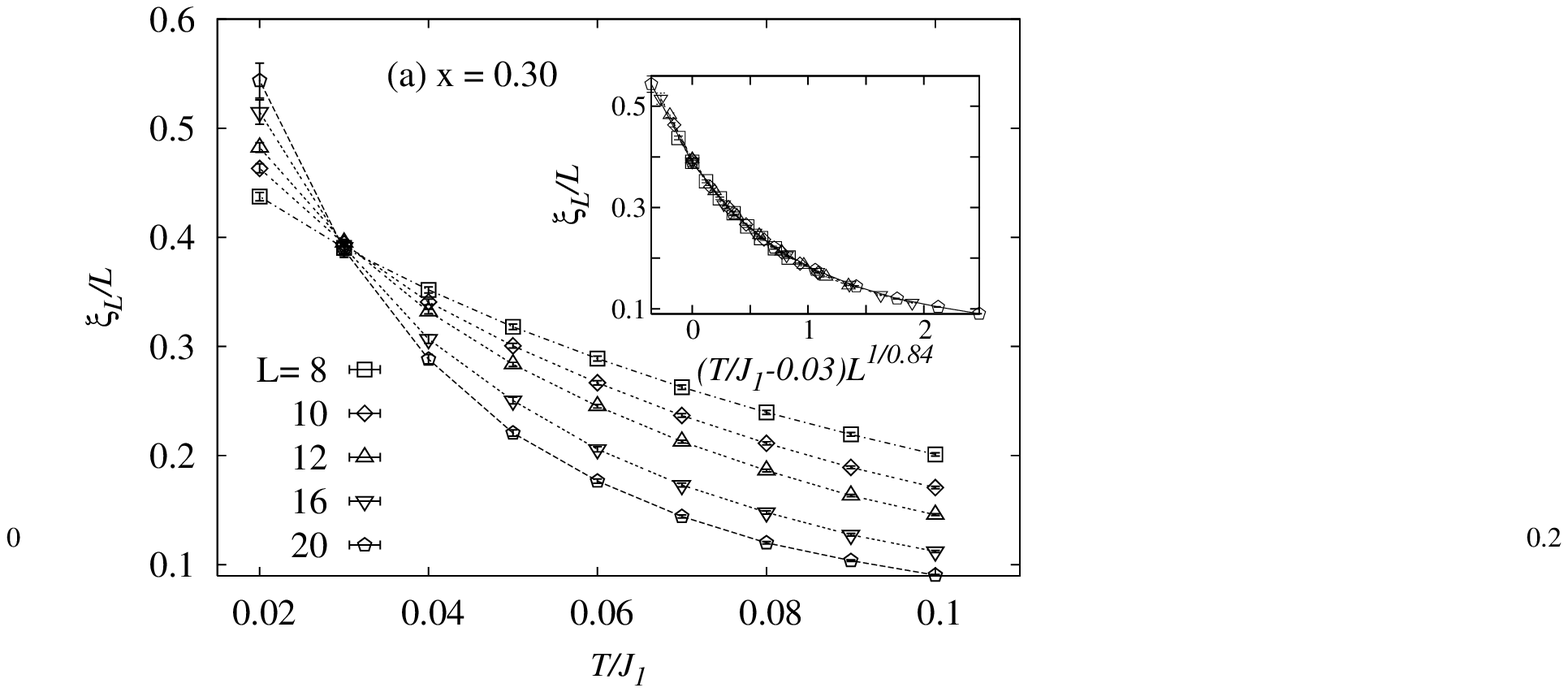}\\
\vspace{-0.2cm}
\includegraphics[width=7.0cm,clip]{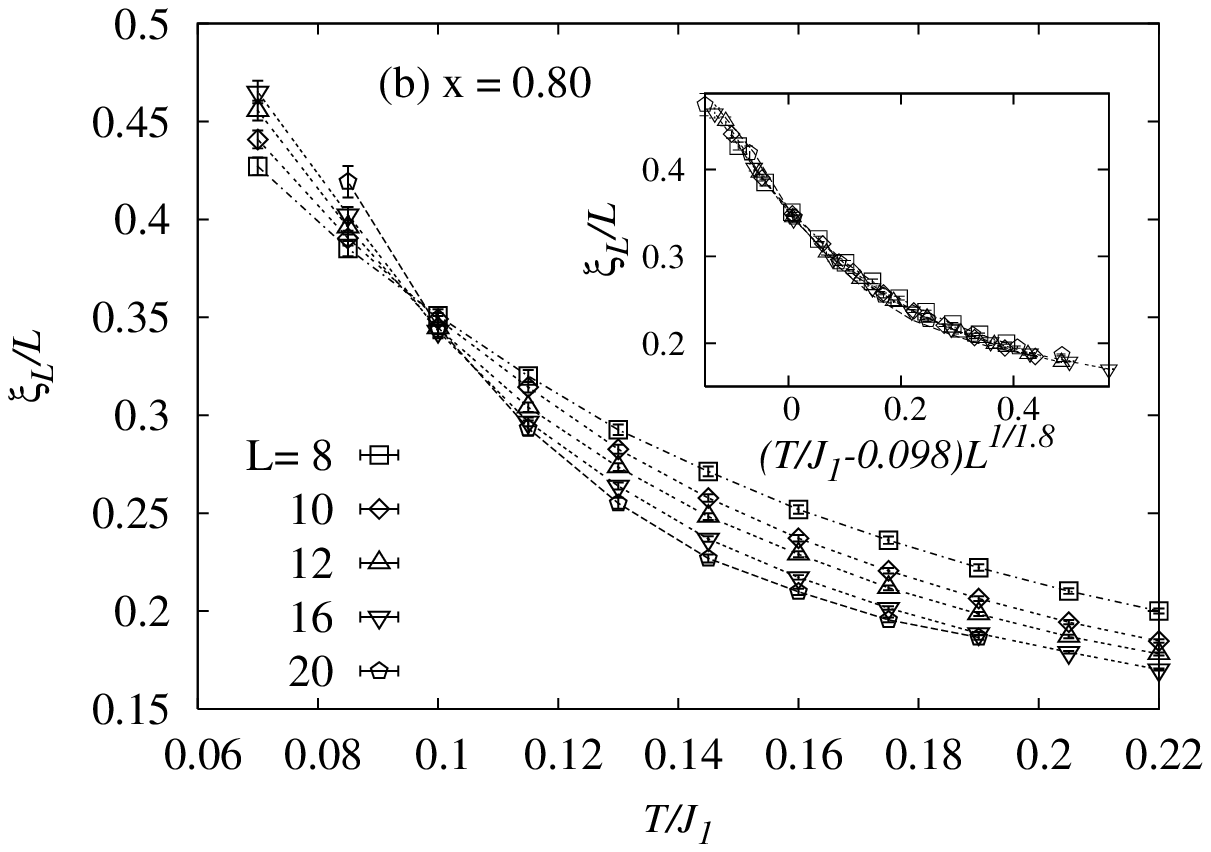}\\
\vspace{-0.2cm}
\includegraphics[width=7.0cm,clip]{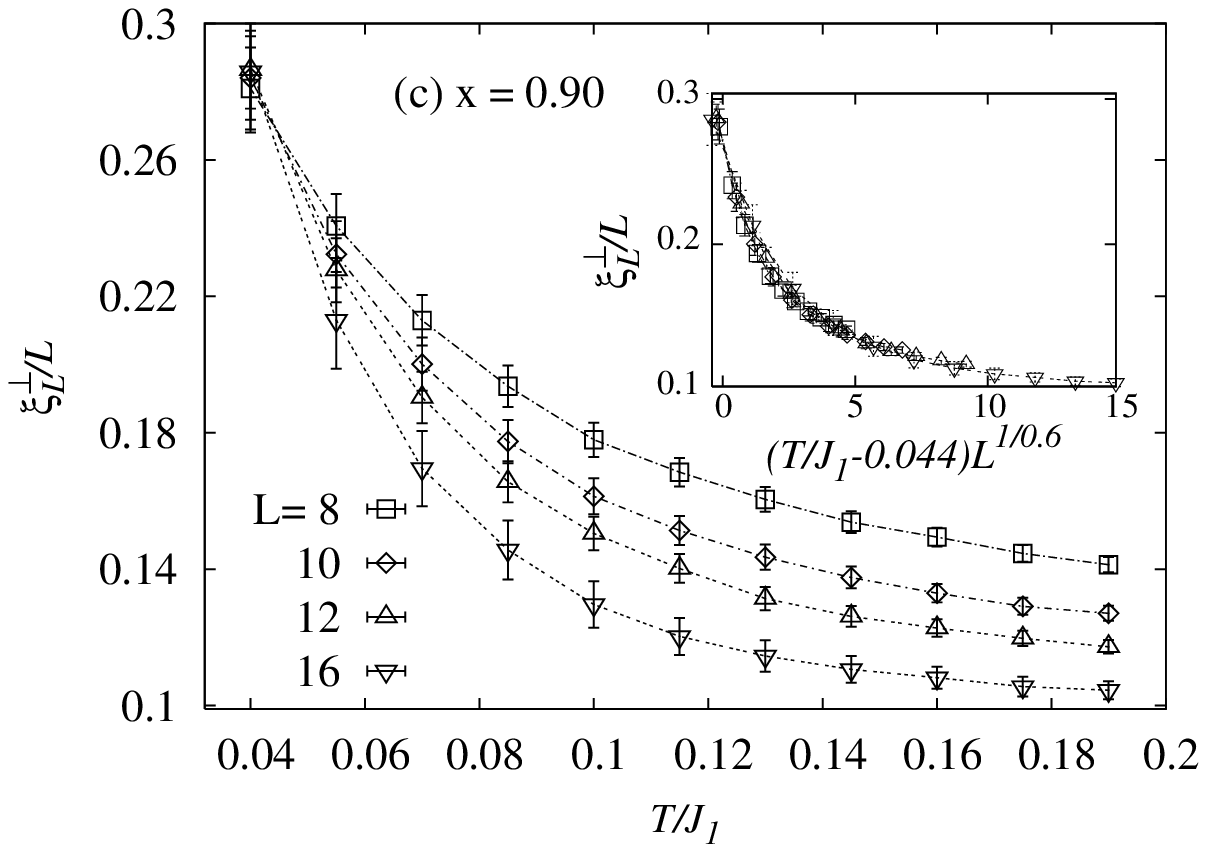}\\
\end{center}
\vspace{-0.4cm}
\caption{
\label{fig:11} 
The SG correlation length $\xi_L$ divided by $L$ at different $x$. 
Insets show typical examples of the scaling plot. 
}
\vspace{-0.4cm}
\end{figure}

Is the SG phase realized at low temperatures? 
A convincing way of examining the SG phase transition is a finite 
size scaling analysis of the correlation length, $\xi_L$, of 
different sizes $L$\cite{Ballesteros, Lee}. 
Data for the dimensionless ratio $\xi_L/L$ are expected to intersect at 
the SG transition temperature of $T_{\rm SG}$. 
Here we consider the correlation length of the SG component of the spin, 
i.e., $\tilde{\bm S}_i (\equiv {\bm S}_i - {\bm m})$ with ${\bm m}$ as
the ferromagnetic component of ${\bm m} = \sum_ix_i{\bm S}_i/(xN)$. 
We perform a cooling simulation of a two-replica system with 
$\{{\bm S}_i\}$ and $\{{\bm T}_i\}$\cite{Bhatt}. 
The SG order parameter, generalized to wave vector ${\bm k}$, 
$q^{\mu\nu}(\bm{k})$, is defined as
\begin{eqnarray}
 q^{\mu\nu}(\bm{k}) = 
\frac{1}{xN}\sum_{i}\tilde{S}_i^{\mu}\tilde{T}_i^{\nu}e^{i{\bm k}{\bm R}_i}, 
\end{eqnarray}
where $\mu, \nu = x, y, z$. From this,the wave vector dependent 
SG susceptibility $\chi_{\rm SG}({\bm k})$ is determinate as
\begin{eqnarray}
 \chi_{\rm SG}({\bm k})=xN\sum_{\mu,\nu}
                      [\langle|q^{\mu\nu}(\bm{k})|^2\rangle]. 
\end{eqnarray}
The SG correlation length can then be calculated from
\begin{eqnarray}
 \xi_L = \frac{1}{2\sin(k_{\rm min}/2)}
         (\frac{\chi_{SG}(0)}{\chi_{SG}({\bm k}_{\rm min})} - 1)^{1/2},
\end{eqnarray}
where ${\bm k}_{\rm min} = (2\pi/L,0,0)$. 
It is to be noted that, in the FM phase (${\bm m} \neq 0$ for 
$L \rightarrow \infty$), the FM component will interfere 
with the development of the correlation length of the SG component 
$\tilde{\bm S}_i$. 
Then in that case 
we consider the transverse components $\tilde{\bm S}_i^{\perp} 
(\equiv (\tilde{\bm S}_i\times {\bm m})\times{\bm m})$ 
in eq. (13) instead of $\tilde{\bm S}_i$. 
The correlation length obtained using $\tilde{\bm S}_i^{\perp}$ is denoted 
as $\xi_L^{\perp}$.


We calculate $\xi_L/L$ or $\xi_L^{\perp}/L$ for $0.20 \leq x \leq 0.90$. 
The crosses for different $L$ are found for $0.30 \leq x \leq 0.90$. 
Figures 12(a)--12(c) show results of the temperature dependence 
of $\xi_L/L$ for typical $x$. 
Assuming that the SG transition occurs at the crossing temperature, we 
can scale all the data for each $x$ (see insets). 
For $x = 0.20$, the crosses were not visible down to $T/J_1 = 0.02$. 
However, we can scale all the data assuming a finite transition temperature 
of $T_{\rm SG}/J_1 \sim 0.01$. Thereby, we infer that the SG transition 
occurs for $0.20 \lesssim x \lesssim 0.90$. 
This finding is compatible with the argument in the previous section that 
$\theta^{\rm SG} > 0$ for $0.20 \lesssim x \lesssim 0.90$. 


It is noteworthy that the SG phase transition for ${\bm m} \neq 0$ is one 
in which the transverse spin components $\{\tilde{\bm S}_i^{\perp}\}$ order. 
Therefore we identify this phase transition as a Gabay and Toulouse (GT) 
transition\cite{GT} and the low temperature phase as a mixed (M) phase 
of the FM and a transverse SG.
It is also noteworthy that, for $x = 0.79$ and $x = 0.80$, we estimate 
respectively $T_{\rm SG}/J_1 = 0.10 \pm 0.01$ 
and $T_{\rm SG}/J_1 = 0.098 \pm 0.005$, whereas respectively 
$T_{\rm R}/J_1 = 0.15 \pm 0.01$ 
and $T_{\rm R}/J_1 = 0.125 \pm 0.005$\cite{Comm_Error}.  
These facts suggest that, as the temperature is decreased, 
the SG transition occurs after the disappearance of 
the FM phase ($T_{\rm SG} < T_{\rm R}$). 
The difference in transition temperatures of 
$T_{\rm INV}( \equiv T_{\rm R})$ and $T_{\rm SG}$ were reported 
in Fe$_{0.7}$Al$_{0.3}$\cite{Motoya}. 
However, further studies are necessary to resolve this point 
because the treated lattices of $L \leq 20$ for estimating $T_{\rm SG}$ 
are not sufficiently large.


\section{Phase diagram}

\begin{figure}[tb]
\includegraphics[width=7.0cm,clip]{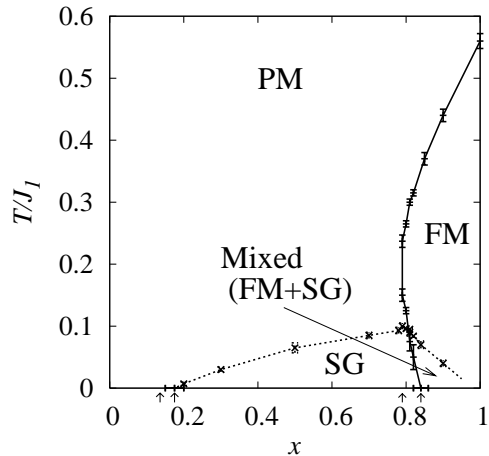}
\vspace{-0.4cm}
\caption{\label{fig:12}
The phase diagram of the dilute Heisenberg model. Four arrows indicate, 
from the left to the right, the percolation threshold $x_{\rm p}$, 
the lower threshold of the SG phase $x_{\rm SG}$, the threshold of the 
ferromagnetic phase at finite temperatures $x_{\rm FT}$, and 
the ferromagnetic threshold at $T = 0$, $x_{\rm F}$. 
}
\end{figure}

\begin{figure*}[tb]
\vspace{-0.0cm}
\hspace{-2.5cm} (a) $x = 0.70$
\hspace{3.0cm}  (b) $x = 0.80$
\hspace{3.0cm}  (c) $x = 0.85$\\
\vspace{0.3cm}
\hspace{0.0cm}\includegraphics[width=3.8cm,clip]{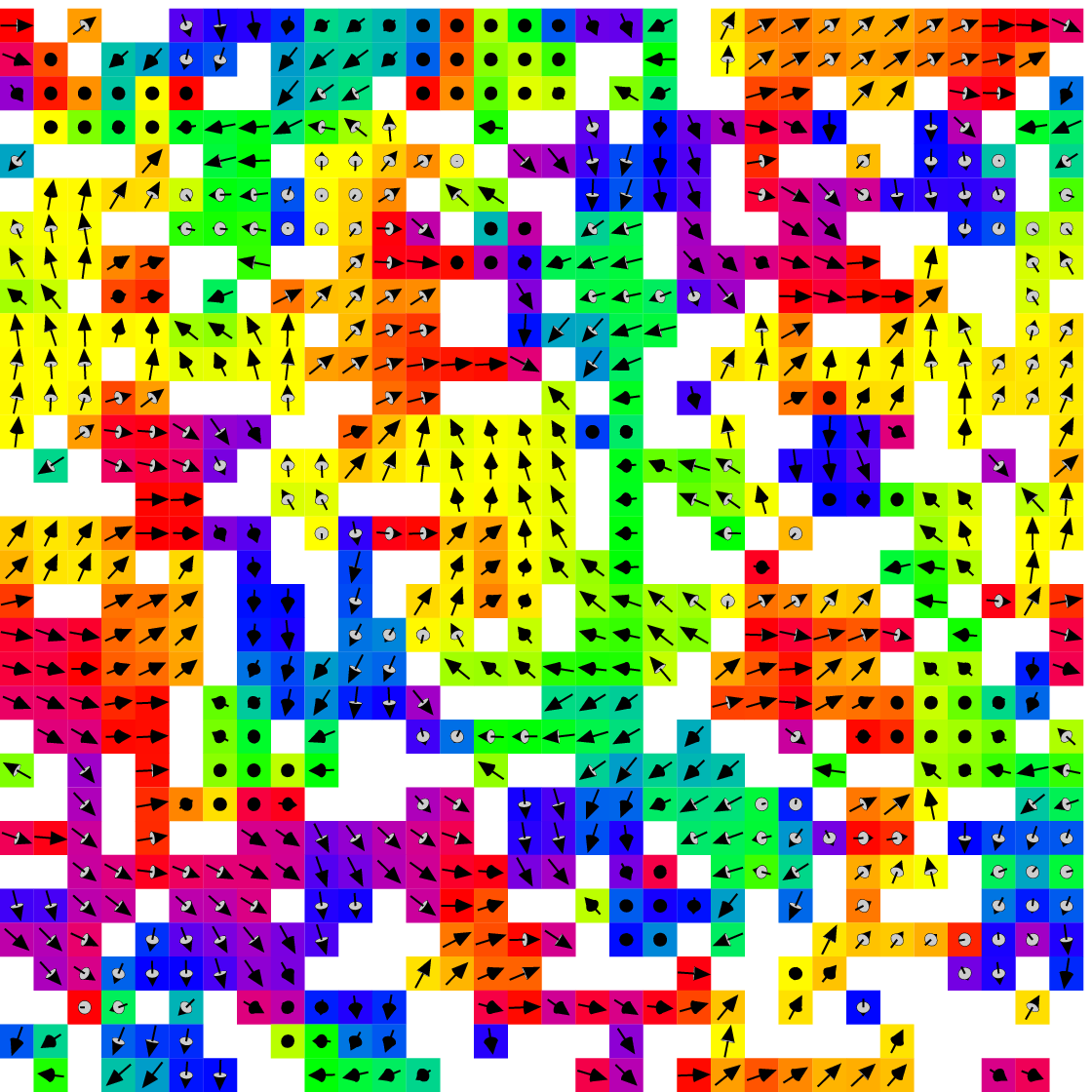}
\hspace{1.0cm}\includegraphics[width=3.8cm,clip]{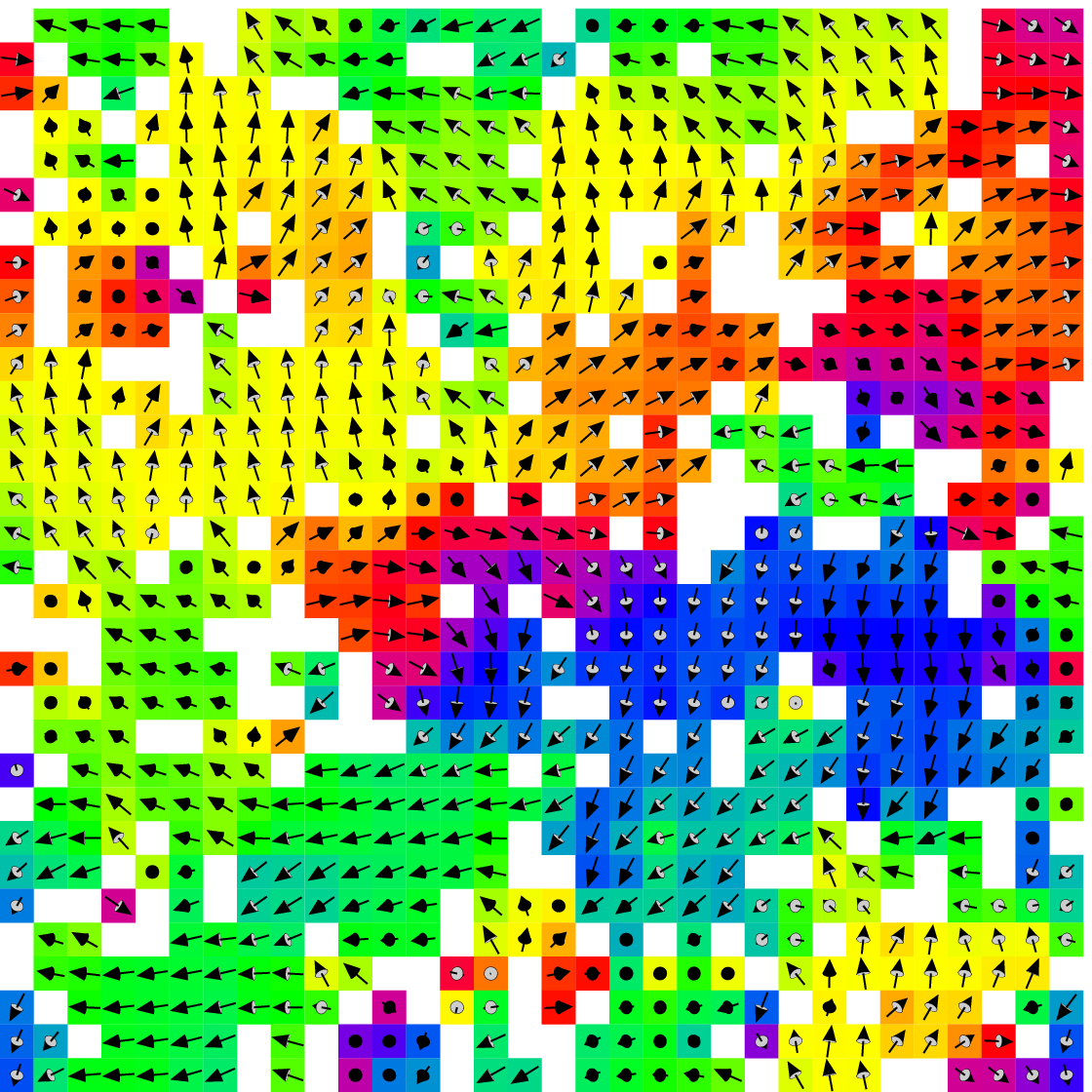}
\hspace{1.0cm}\includegraphics[width=3.8cm,clip]{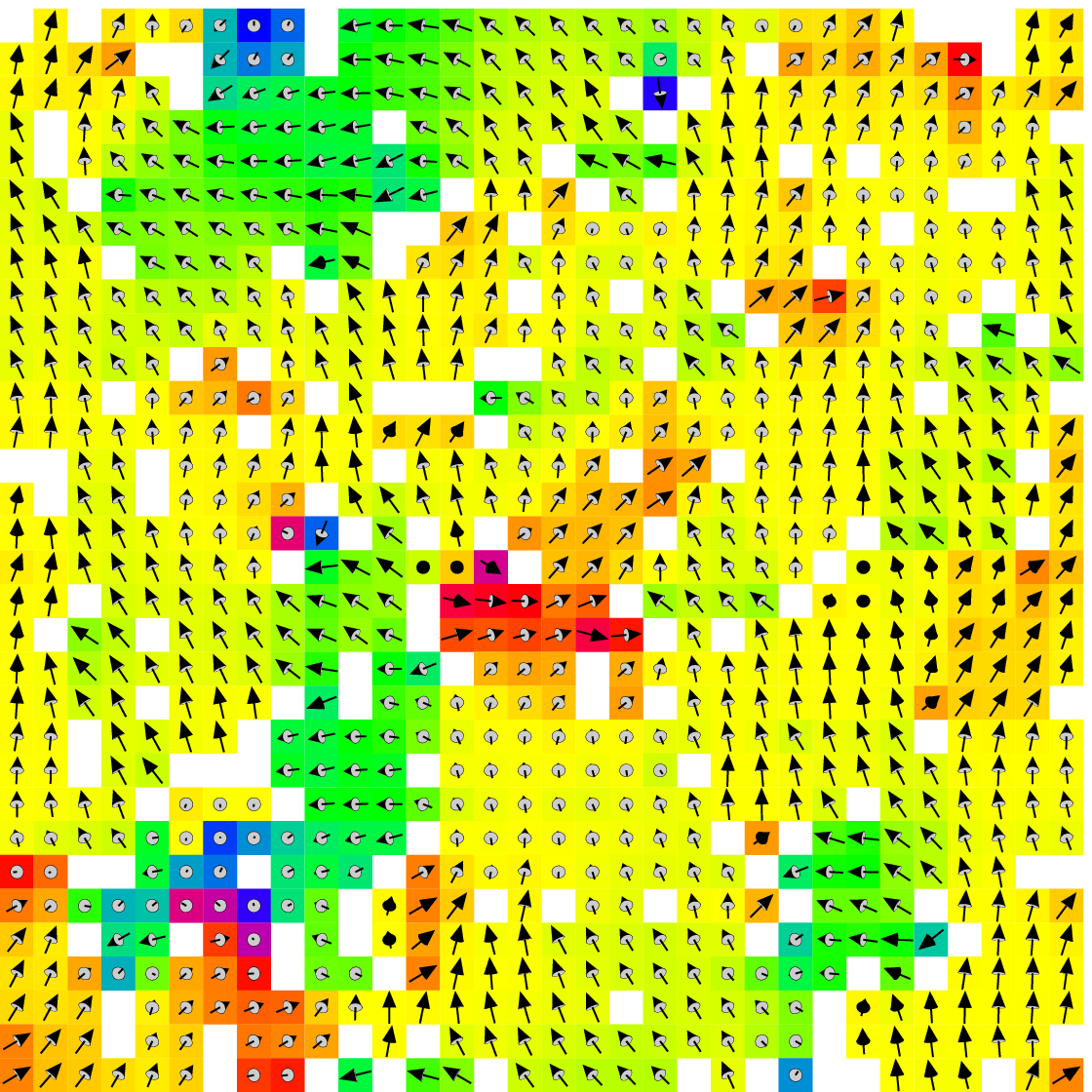}
\hspace{0.3cm}\includegraphics[width=1.5cm]{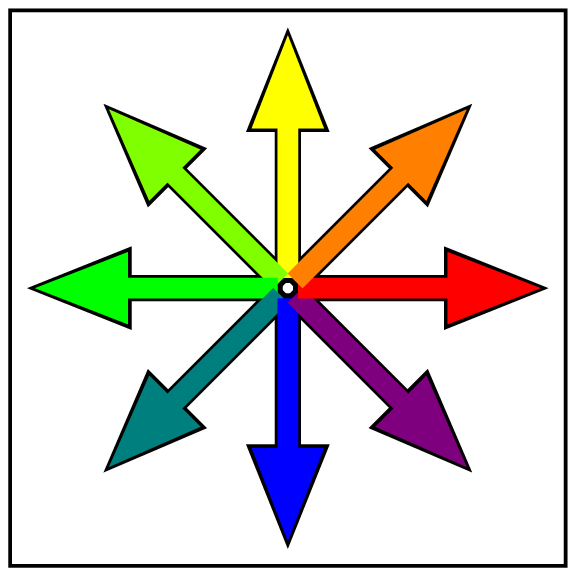}
\caption{ \label{fig:13}
(Color online) Spin structures of the model for different $x$ at 
$T/J_1 = 0.04$ on a plane of the $32 \times 32 \times 32$ lattice. 
Spins represented here are those averaged over 10000 MCS. 
The positions of the non-magnetic atoms are represented in white.
}
\end{figure*}

Figure 13 shows the phase diagram of the model obtained in this
study. It is shared by four phases: (i) the PM phase, (ii) the FM phase, 
(iii) the SG phase, and (iv) the M phase. 
A point that demands re-emphasis is that, just below the $T = 0$ phase 
boundary between the SG phase and the M phase 
$(x_{\rm FT} < x < x_{\rm F})$, the RSG transition is found. 
This phase diagram is analogous with those observed in dilute ferromagnets 
Fe$_x$Au$_{1-x}$\cite{Coles} and Eu$_x$Sr$_{1-x}$S\cite{Maletta1,Maletta2}. 
In particular, the occurrence of the mixed phase was reported in 
Fe$_x$Au$_{1-x}$. 


We examine the low temperature spin structure. 
Figures 14(a) and 14(b) represent the spin structure 
in the SG phase ($x < x_{\rm F})$. 
We can see that the system breaks up to yield ferromagnetic clusters. 
In particular, for $x \lesssim x_{\rm F}$ (Fig. 14(b)), the cluster size 
is remarkable.  
Therefore the SG phase for $x \lesssim x_{\rm F}$ is characterized by 
ferromagnetic clusters with different spin directions. 
Figure 14(c) represents the spin structure in the M phase ($x > x_{\rm F})$. 
We can see that a ferromagnetic spin correlation extends over the 
lattice. There are ferromagnetic clusters in places. 
The spin directions of those clusters tilt to different directions. 
That is, as noted in the previous section, the M phase is characterized 
by the coexistance of the ferromagnetic long-range order and the ferromagnetic 
clusters with transverse spin component.  
The occurrence of ferromagnetic clusters at $x \sim x_{\rm F}$ are compatible 
with experimental observations\cite{Coles,Maletta1,Maletta2,Motoya,Yeshurun}.


\section{Conclusion}

This study examined the phase diagram of a dilute ferromagnetic Heisenberg 
model with antiferromagnetic next-nearest-neighbor interactions. 
Results show that the model reproduces experimental phase diagrams 
of dilute ferromagnets. 
Moreover, the model was shown to exhibit reentrant spin glass (RSG) behavior, 
the most important issue. 
Other important issues remain unresolved, especially in the RSG transition. 
Why does the magnetization, which grows at high temperatures, diminish at 
low temperatures? Why does the spin glass phase transition take place 
after the disappearance of the ferromagnetic phase? 
We intend the model presented herein as one means to solve those and other 
remaining problems. 

\bigskip

The authors are indebted to Professor K. Motoya for directing their attention 
to this problem of the RSG transition and for his valuable discussions. 
The authors would like to thank Professor T. Shirakura and 
Professor K. Sasaki for their useful suggestions.
This work was financed by a Grant-in-Aid for Scientific Research 
from the Ministry of Education, Culture, Sports, Science and Technology.




\begin{references}
\bibitem{Model} Coordination numbers of the nearest and the next-nearest 
          neighbor lattice sites are $z_1 = 12$ and $z_2 = 6$ in the fcc 
          lattice, and $z_1 = 6$ and $z_2 = 12$ in the sc lattice. 
          In that case, we choose the smaller ratio of $J_2/J_1 = 0.2$ 
          instead of $J_2/J_1 \sim 0.5$ in Eu$_x$Sr$_{1-x}$S.
\bibitem{Comm_Endoh} It was found that $\Delta E_L(\phi)$ of the $\pm J$ 
          Heisenberg model is expressed as a product of decoupled two 
          functions: $\Delta E_L(\phi) = F(\phi)L^{\theta}$\cite{Endoh1}. 
          So we suppose that it is also ture in any Heisenberg model and 
          examine it in a single $\phi$.
\bibitem{Comm_finite} If we estimate the value of $\theta^{\rm SG}$ using raw 
        data, we get an extraordinarily large value of $\theta^{\rm SG} > 1$.
\bibitem{Comm_gL} 
         Values of $g_L$ depend strongly on the sample number $N_s$. 
         Figures 11(a)--11(d) show $g_L$ when  $g_L$ for $N_s$ 
         is not changed considerably from that for $N_s/2$.
\bibitem{Comm_Error} 
        The error bars for $T_{\rm SG}$ are estimated from the scaling plot; 
        those for $T_{\rm R}$ are estimated from scattering of the crossing 
        temperatures.
\end{references}
\end{document}